\documentclass[singlecol]{epl2} 
\usepackage{amssymb,bm}
\usepackage{amsmath,eucal,ulem}
\usepackage{graphicx}
\usepackage{comment}

\title{Adaptation and irreversibility in microevolution.}
\shorttitle{Adaptation and irreversibility in microevolution.}
         \author{Stefano Bo\inst{1}\and Andrea Mazzolini\inst{1} and Antonio Celani\inst{2}}
         \shortauthor{S. Bo \etal}

         \institute{ 
         \inst{1}             Physics Department and INFN, University of Turin, via P. Giuria 1, I-10125 Turin, Italy.\\

           \inst{2}The Abdus Salam International Centre for Theoretical Physics (ICTP), Strada Costiera 11, I-34014 - Trieste, Italy.\\
          
           }
           

\begin{document}
\abstract{Within the framework of population genetics we consider the evolution of an asexual haploid
population  under the effect of a rapidly varying natural selection (microevolution). We focus on the case in which the environment
exerting selection changes stochastically. We derive the effective genotype and fitness dynamics on the slower time-scales at which
the relevant genetic modifications take place. We find that, despite the fast environmental switches, the population manages to adapt on 
the fast time-scales yielding a finite positive contribution 
to the fitness. However, such contribution is balanced by the continuous loss in fitness due to the varying selection so that
the statistics of the global fitness can be described neglecting the details of the fast 
environmental process. 
The occurrence of adaptation on fast time-scales would be undetectable if one were to consider only the effective genotype and fitness dynamics on the slow
time-scales. We therefore propose an experimental observable to detect it.
}
\maketitle
\section{Introduction}
\begin{figure*}[t!]

\includegraphics[width=0.48\columnwidth]{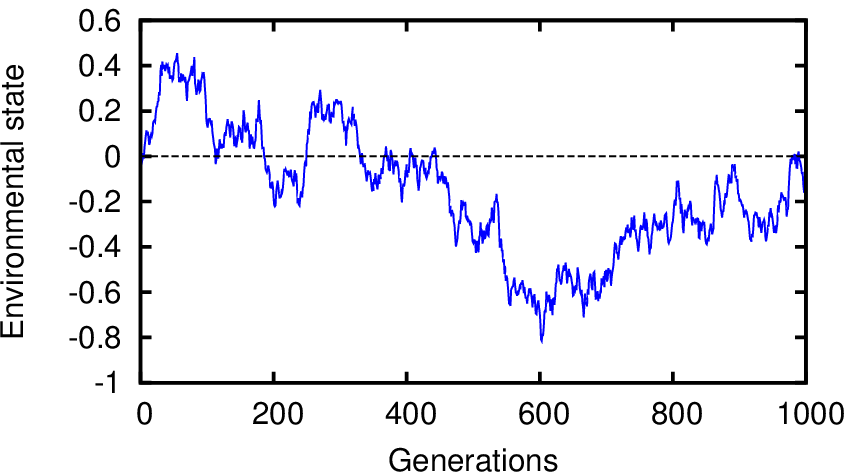}
\includegraphics[width=0.48\columnwidth]{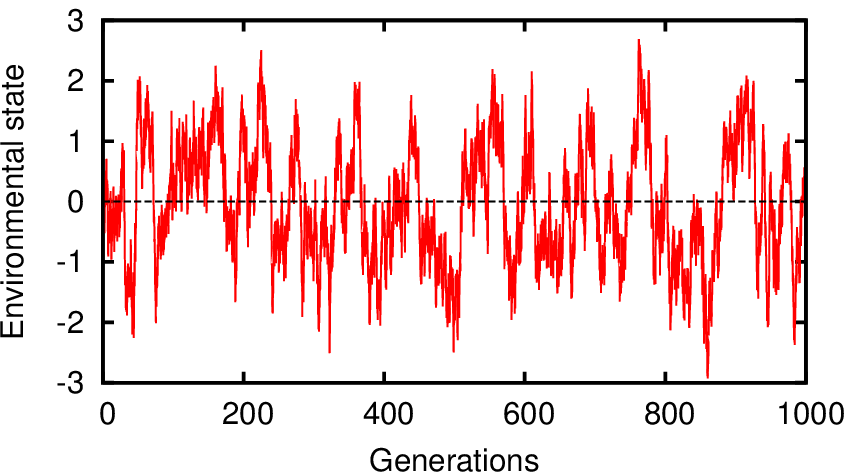}
\includegraphics[width=0.48\columnwidth]{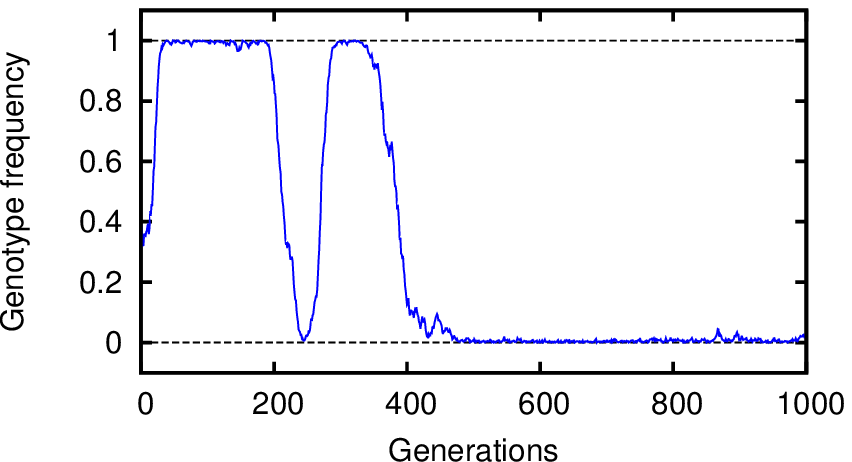}
\includegraphics[width=0.51\columnwidth]{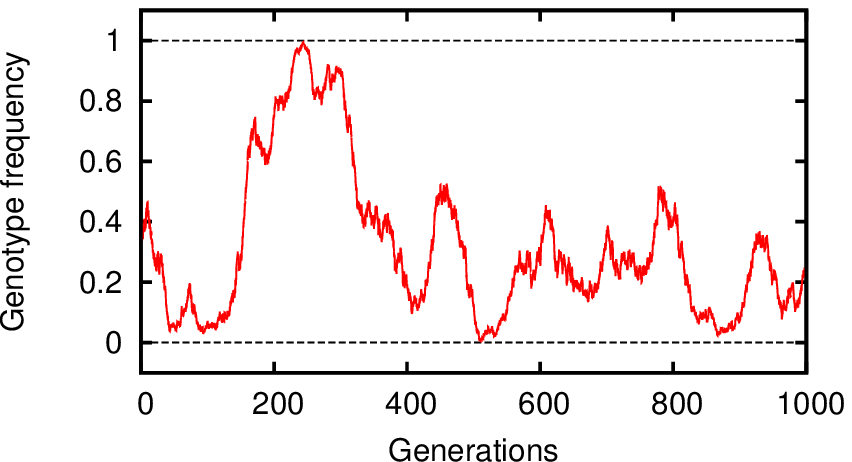}
\label{fig:micro}
\caption{
Simulation of the two genotypes dynamics obeying Eq.~(\ref{eq:ko}) coupled with an environment following an Ornstein-Uhlenbeck process (\ref{eq:env_ex}). 
The left panels refer to environment changes with a correlation time:
$T_e = 1000 \text{ gen}$, which is of the same order of the typical time of genotype variation. The right panels
describe the microevolutionary regime with environmental time correlations of $T_e = 5\text{ gen}$. For the chosen parameters significant changes 
in the genotype frequency occur after typical times of $100-1000$ generations. }
\end{figure*}

Population genetics studies the evolutionary process of a population under forces such as mutations,
 Darwinian selection and the random genetic drift 
 (see \cite{hartl_clark,gillespie,blythe07} for a general exposition).
A key intuition for modeling the effects of natural selection is the concept of fitness
 of a population in a certain 
environment  which describes the global ability of the population to reproduce and survive~\cite{haldane27}. 
The different fitness of various genotypes can be effectively visualized in terms
of a fitness landscape~\cite{wright} which is usually "climbed" during the course of evolution.
Such climbing is referred to as adaptation.
Fisher's fundamental theorem of natural selection~\cite{fisher}
states that when evolution is subject only to natural selection
in a constant environment, the fitness of a population increases
at a positive rate equal to the variance of the population.
Indeed, 
natural selection, by its very definition, is the force that favors fitter individuals,
pushing towards the genotype configuration which maximizes the global fitness.
When mutations and random drift are relevant, the stochastic nature of evolution
emerges and adaptation becomes a more complex phenomenon. 
One of the most widely accepted model for the stochastic description of the evolutionary process is the Kimura-Ohta
 equation~\cite{kimura64} that we will describe in detail in Eq.~(\ref{eq:ko}) below.
In most natural cases the environment in which the population evolves changes in time
(see for example Refs. \cite{mustonen08,mustonen10,peischl12,otinowski11,tanasenicola12,bell10,rivoire14,held14})
and genotypes that were fit under the initial condition may successively be unfavored by selection.
 Consider for instance the case of a population of bacteria shaped by natural selection to
 metabolize a certain nutrient. If such nutrient is gradually replaced by a different 
 one  the metabolism of the bacteria (optimized for the first one) will now be less efficient and the global fitness of the bacterial 
population will decrease. At the same time adaptation to this new environment
will start taking place providing a fitness increase.
We can then represent the fitness dynamics as the result of two effects: 
adaptation driven by 
natural selection (analogous to the term appearing in Fisher's theorem \cite{price72}) and fitness changes due to environmental variations:
\begin{equation}\label{eq:fit_gen}
    \frac{d}{dt} \text{Fitness} = \text{Adaptation} + \text{Environmental changes}
\end{equation}
We will consider the explicit expression for the fitness variation and its quantitative interpretation below (see eq.~\ref{eq:dfit}).

The changes brought about by the variability of the environment give a negative contribution
countered by adaptation, 
which is, in general, positive.

In the present work we consider the case in which the time scales of the environmental changes are much faster than the
 typical time of evolution of genotype frequencies due to evolutionary forces.
 Such scenario is referred to as the
microevolutionary limit (see Fig.1). 
For the example of the bacteria culture in different growth media, microevolution
corresponds to replacing the nutrients at a much faster rate than the changes in the genetic structure of the population of bacteria
 driven by evolution.
In particular we focus on the case in which the dynamics of the environment is a stochastic process itself (as done also in
Refs.\cite{mustonen08,peischl12}).
We are then 
considering a population in a rapidly fluctuating environment thereby subject to erratic selection.
 In order to understand our main result let us notice that if one studies the evolution of the population on long timescales it may seem neutral
because the random contributions of the selection exerted by the succession of environments tend to cancel 
off\footnote{In the discussion we consider the general case in which the stochastic
selection may have a non-zero average.}.
However, this apparent
neutrality does not mean that the population has not been subject to selection and that it
has not adapted to the environments it has encountered. In the following we  
show how to detect and account for this hidden adaptation.

We also study the dynamics of the global fitness of the population in 
 the case in which the stochastic dynamics of the environment reaches a statistically steady state\footnote{A statistical steady state of the environment
 implies that
the state of the environment is not fixed but continually changes in a stochastic manner. However, the probability distribution of the environment is constant.}. 
We expect that, given the statistical stationarity of the environment, after an initial adaptation, the average of the global
fitness is constant in time.
To check this intuition and to investigate in depth the issue 
we make use of asymptotics techniques \cite{pavliotis} to derive the effective evolution
 taking place at the relevant 
slow time scales (much longer than the correlation time of the environment).
We find that the genotype dynamics becomes independent of the 
specific environmental state and is governed by the average of
the selection coefficient.
As expected, the global fitness is constant but not in a "static" way.
It is rather the sum of
 two opposite finite terms that compensate each other:
positive adaptation and the change of fitness due to the variability of the environment
which gives negative contributions.
The positive term  accounts for 
the adaptation that has taken place on the fast timescales. It can be traced back to the fact that, even when the typical fluctuation time
 during which the environment favors a specific genotype 
tends to zero, the population undergoes an infinitesimal adaptation,
 which is of the order of the fluctuation time. The number of 
such infinitesimal positive contributions in a unit of time increases proportionally to the 
inverse of the typical time of fluctuations, so that they constitute a finite additional term. 
Such mechanism
is analogous to the one recently
found for entropy production in stochastic systems with fast and slow timescales \cite{celani12,bo14}.
We can give an intuitive picture for the balance of contributions leading to a constant fitness by considering a runner on a treadmill.
 The speed of the treadmill represents the loss in fitness due to the fast environmental changes.
 In order to maintain a fixed position (i.e. constant fitness) the runner has to run (i.e adaptation to the environment)
 at the same and opposite speed of the
 treadmill.

In the following sections we will detail the formalism underlying our discussion.
We will present our results using the mathematical model of Kimura-Ohta
and give a general example.
To conclude we will suggest an observable which can provide experimental 
measures of the found adaptive behavior.


\section{Fitness and adaptation in a fluctuating environment}
For the sake of clarity we illustrate the simplest situation considering 
a haploid population which can switch between two genotypes: $A$ and $B$.
We denote the frequency of $A$ as $x$ 
so that the Kimura-Ohta equation reads:
\begin{equation}\label{eq:ko}
\frac{\partial P}{\partial t} =- \frac{\partial}{\partial x}\left[ (m + gs) P\right] 
+ 
 \frac{1}{2N} \frac{\partial^2}{\partial x^2} \left[g P \right]
 \end{equation}
 
where time is expressed in generation units. 
 $P(x,t)$ is the probability distribution that, at time $t$, the frequency of genotype $A$ is $x$.
  $m(x) = -\mu x + \nu (1-x)$ is the mutation coefficient, $\mu$ and $\nu$ are the mutation rates.
   $s(x,y) = f_A(x,y) - f_B(x,y)$ is the selection coefficient, which describes the effect of natural selection
 and depends on the
  environmental state. $f_{A/B}(x,y)$ is the fitness of the genotype $A/B$. Note that we are considering the fitness to depend 
  on the environmental state which
  we denote as $y$. 
      $N$ is the effective population size and $g(x) = x (1-x)$.
The global fitness of the population, $F(x,y)$, is defined by the following relation:
\begin{equation}\label{eq:s}
  s(x,y) = \frac{\partial F(x,y)}{\partial x}
\end{equation}
that is, the function that the natural selection tends to maximize.
This identification is valid when the selective advantage can
be expressed in terms of  the gradient of a fitness landscape.
We are therefore
not taking into account the more involved settings of cyclic selective advantages
discussed for example in Refs. \cite{kerr02,sinervo06,frey}.
When the selection coefficient does not depend on the genotype frequencies
(frequency independent selection) 
 the global
 fitness corresponds (up to an additive constant) to the average population fitness:
 $\langle f \rangle = f_A(x,y)x + f_B(x,y)(1-x)$ (see Ref. \cite{mustonen10}
for a detailed discussion).
We will consider such instance in the example below.
On the contrary, when selection depends on the genotype frequency the global fitness
 differs from the average population fitness.
 In general, $F$ can be interpreted as the fitness landscape
i.e.  the quantity that Darwinian selection tends to maximize.
Indeed, its gradient is what sets the direction 
in which natural selection drives evolution (see eq.~\ref{eq:s}). 
In other words, natural selection pushes towards a set of genotype frequencies with a higher $F$.
To appreciate this property let us consider  an example involving a frequency dependent selection
highlighting the difference with the average fitness.
Let us take the first genotype fitness to be $f_A(x) = \alpha 
+ 1 - x$, and the second one $f_B(x) = 1 - x$ with $0<\alpha <1$. 
In this case, the first genotype is favored by natural selection, $s = f_A - f_B = \alpha > 0$, and increases in number through generations.
The global fitness $F(x) = \alpha x$ consistently grows along this adaptive process in which  $x$ increases. On the contrary, the average
fitness, $\langle f \rangle = x f_A(x) + (1-x)f_B(x) = (\alpha - 1)x + 1$,  decreases with the increase of $x$ and is therefore not suitable 
for characterizing the adaptation which is taking place. 

Let us now study in detail the infinitesimal variation of the global fitness introduced in eq.~(\ref{eq:fit_gen}).
 It can be written as
\begin{eqnarray}\label{eq:dfit}
  d F(X_t,Y_t) =& \underbrace{s(X_t,Y_t) \circ dX_t}_{\mbox{adaptation}}+\underbrace{\frac{\partial F}{\partial y} \circ dY_t}_
  {\mbox{env. changes}} 
\end{eqnarray}
where $X_t$ and $Y_t$ are, respectively, the first genotype frequency and the environmental 
state at time $t$, and "$\circ$" is a Stratonovich product \footnote{For a generic function $f(X_t)$
the Stratonovich product implies the midpoint discretization $f(X_t)\circ dX_t\equiv\frac{f(X_t)+f(X_{t+\tau})}{2}\left(X_{t+\tau}-X_t \right)$.}.
We remark that the $dX_t$ is given by the genotype evolution (following eq.~\ref{eq:ko}) and includes the effects of mutations and genetic drift.
The first term on the right hand side 
accounts for the changes in fitness caused by changes in the genotype frequencies. It coincides with the definition of the fitness flux at time $t$ introduced
 in Ref.\cite{mustonen10}
\begin{equation}\label{eq:ff}
  d\phi(t) = s(X_t,Y_t) \circ dX_t=\frac{\partial F}{\partial x}\circ dX_t
\end{equation}
which measures the adaptation driven by natural selection following the intuition used by Fisher in the derivation of his fundamental
theorem~\cite{price72}. 
To probe this interpretation consider the case in which the first genotype has a higher fitness
so that the selection coefficient is positive. 
If the first genotype frequency grows, the population is adapting to the environment 
 and we see that this corresponds to a positive  fitness flux.
 A a dis-adaptive behavior would imply a decrease in the first genotype frequency and consequently a negative fitness flux. 
The second term on the right hand side of Eq.~(\ref{eq:dfit}) accounts for
 the variation of fitness at fixed genotype frequency due to changes in the environment.
If the population is well adapted these changes usually reduce the global fitness.
The adaptation of a population to an environment clearly shows a temporal
direction and is therefore an irreversible process.
For general stochastic dynamics irreversibility is measured
in terms of entropy production \cite{chetrite08,seifert12}.
The authors of Ref. \cite{mustonen10}
highlighted the connection
between population genetics and out-of-equilibrium stochastic systems
by showing that the fitness flux 
corresponds to the entropy production in the environment of the stochastic dynamics
describing the genotype evolution.
We provide the details in the supplemental material. 
Such analogy allowed the authors of Ref. \cite{mustonen10}
to derive a sort of second law of thermodynamics for population genetics which they dubbed
fitness flux theorem.
Just as entropy production, the fitness flux at the stationary state is,
 on average, always positive $\langle d\phi(t) \rangle \ge 0$ . 
This implies that the evolutionary process is adaptive and irreversible.
 The equality holds if and only if the system is at equilibrium.
 In this case the system is reversible and the population is not adapting.
\section{The microevolutionary limit}

\begin{figure*}[t!]
\includegraphics[width=0.5\columnwidth]{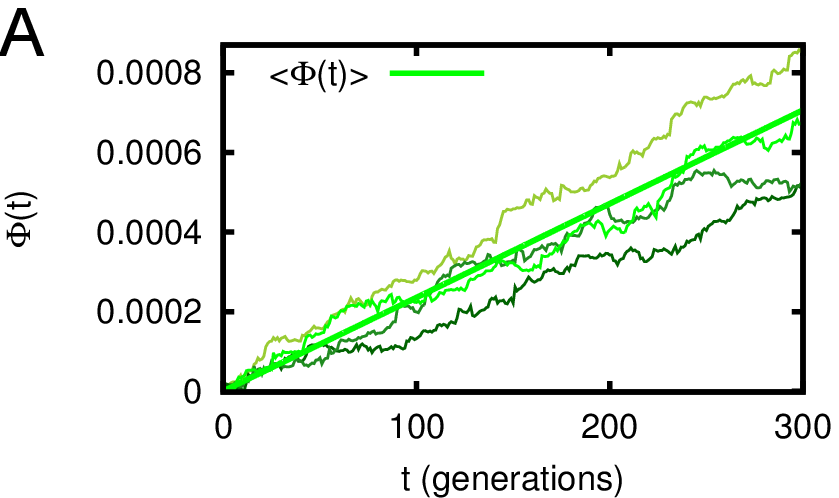}
\includegraphics[width=0.5\columnwidth]{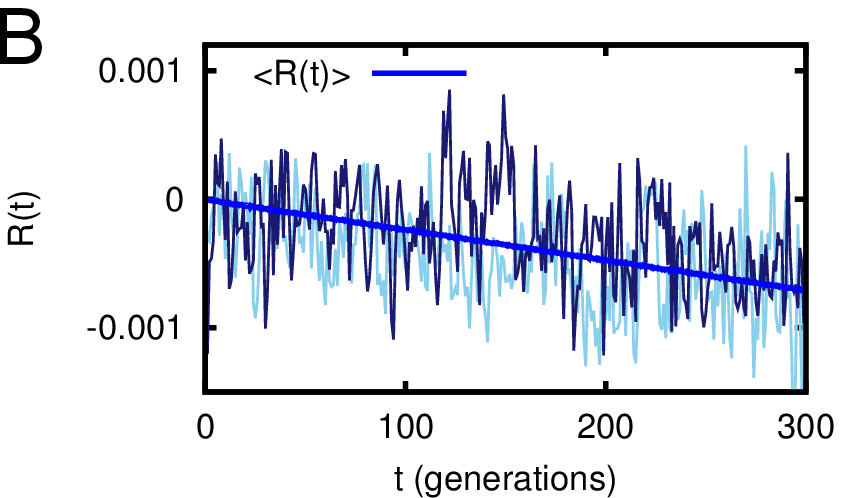}
\includegraphics[width=0.5\columnwidth]{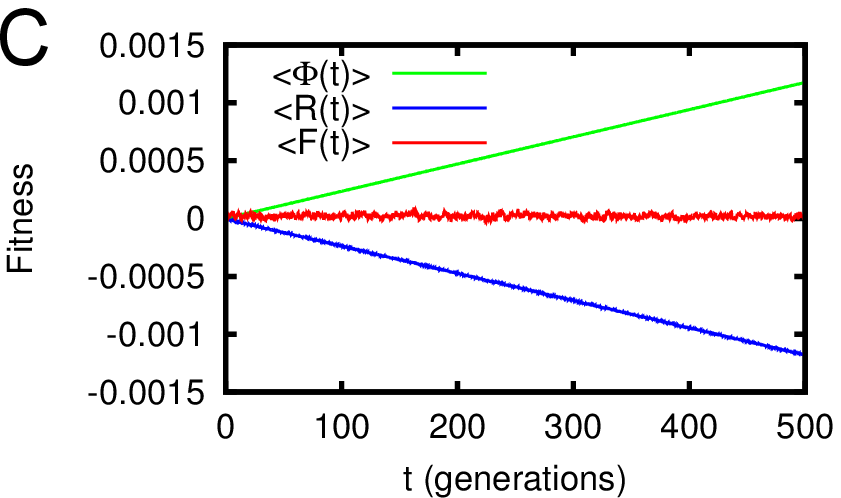}
\includegraphics[width=0.5\columnwidth]{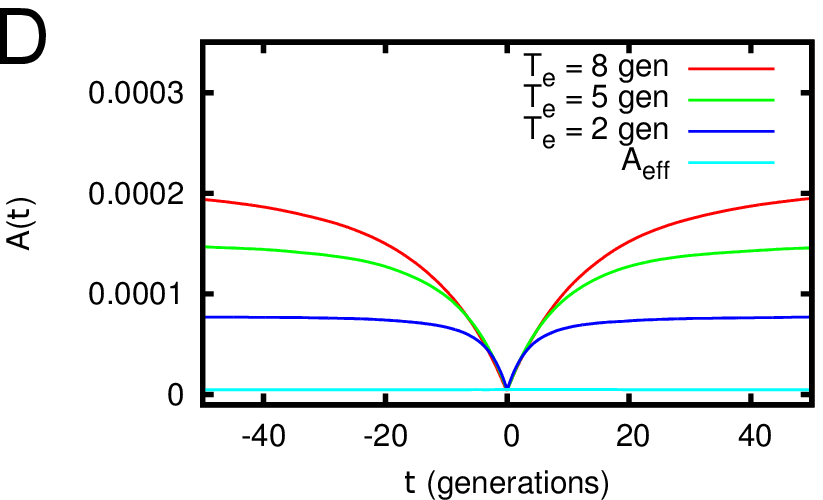}
\caption{Computer simulation of an evolutionary process 
in microevolution at the stationary state. 
The population evolves following the Kimura-Ohta equation (\ref{eq:ko}) 
for two genotypes under a very fast fluctuating environment obeying Eq.~(\ref{eq:env_ex}). 
 The parameters of the process are: $k=1$, $D=0.01$, $N = 10^{4}$, $\mu=\nu = 10^{-3}$, $\sigma = 0.1$. 
Panel A shows four trajectories of the cumulative fitness flux and its average value with 
a slope equal to the average fitness flux.
Panel B depicts the dynamics of the fitness variation due to the environmental changes $R(t)$. 
The slope of the average is the opposite to the average of the fitness flux, but the statistic of this term is completely
different from the statistic of $\Phi(t)$. 
The average fitness dynamics is shown in panel C and, following Eq.~(\ref{eq:dfit}) shows the compensation
of the terms in panel A and B.
Panel D shows the  observable defined in Eq.~(\ref{eq:obs}) computed for  parameters:
$k=1$, $D=1$, $N = 10^5$, $\mu=\nu = 5\cdot 10^{-4}$, $\sigma = 0.02$, $T_e = k^{-1}$.}
\end{figure*}

In this section we investigate what is the asymptotic behavior of the genotype
and the global fitness 
dynamics in the limit of a very rapidly changing environment. In such microevolutionary settings
 the typical time of environmental changes ($\theta = \epsilon^{-1} t$) is much faster than 
typical time of genotype frequencies changes due to evolutionary forces $t$ (See
the right panels of
 fig.1).
We seek to determine what is the effective behavior of the system on the timescale
of the genotype frequency change $t$.
We obtain our results by means of multiple-scale methods (see Ref.~\cite{pavliotis}) taking the  limit $\epsilon \rightarrow 0$.
\subsection{Effective genotype dynamics}
We find that the evolution of the genetic structure of a population in the microevolutionary limit 
is described by 
an effective Kimura-Ohta  equation (Eq.~\ref{eq:ko}) where the selection coefficient is given by its  
average over the possible environments:
\begin{equation}\label{eq:eff_s}
        \overline{s(x)} = \int dy \hspace{1mm} w(y) s(x,y)
      \end{equation}
      and $w(y)$ is  the equilibrium distribution of possible environments.
The effective Kimura-Ohta equation is then independent of the environmental state.
This can be intuitively understood by considering that when the environment changes very rapidly,
the genotype dynamics does not manage to fully react to the natural selection exerted by the current environmental state
before the environment changes again. 
The dynamics is rather affected by a smeared, net contribution of the succession of environments it encounters.
When the environment is in a statistically stationary state such contribution corresponds to the average
selection of the environments. 
%
%
%

\subsection{Fitness dynamics}
At the steady state
the average global fitness is constant in time.
As mentioned in the introduction we find that in microevolution 
this stationarity is achieved by the compensation of
 two opposite finite contributions:
  \begin{equation}\label{eq:eff_df}
     d \langle F(X_t,Y_t) \rangle = \underbrace{\langle d\phi(t)\rangle}_{>0}
 + \underbrace{\langle\frac{\partial F}{\partial y} \circ dY_t\rangle }_{\langle dR\rangle<0} = 0\;.
  \end{equation}
 

Both terms
can be shown to be constant at the steady state (see supplemental 
material) so that their finite time contributions scale linearly in time as shown in Fig.2.
The adaptive fitness flux term displays
an anomalous behavior analogous to that of the entropy production in stochastic systems
with slow and fast scales \cite{celani12,bo14}.
In the microevolutionary limit its average value is always greater 
than the value that one would obtain by simply using
the definition (\ref{eq:ff}) with
the effective dynamics (\ref{eq:eff_s}).
Namely,
      \begin{equation}\label{eq:eff_ff}
        \langle d\phi(t) \rangle = \left\langle \underbrace{ \overline{s(X_t)} \circ dX_t}_{d\phi_{\mathit{eff}}} + 
       \underbrace{g(X_t) \left(\overline{s(X_t)^2}-\overline{s(X_t)}^2\right)dt}_{d\phi_{\mathit{anom}}} \right\rangle 
      \end{equation}
      where the additional contribution is clearly always positive.
In order to understand the implications of this result, consider the  case 
 in which the average selection vanishes i.e. no specific genotype
 is favored in the long term. In this case, the effective Kimura-Ohta
 equation is not subject to natural selection and we would expect no adaptation.
%
%
However, if we compute the limiting value of fitness flux 
we obtain a finite positive value (see Fig.2) testifying that
adaptation is continuously taking place contrasting the fitness loss
due to the fluctuating environment. 

%

\subsection{An experimental observable to probe adaptation in microevolution}
As we have shown, in microevolution 
the trajectories of the genotype frequency follow an effective dynamics which is independent of the environmental state.
Therefore if we want to determine whether the population evolves in a
microevolutionary regime presenting an anomalous adaptation or in a scenario with a fixed
environmental state and a selection coefficient equal to the effective coefficient it is not sufficient to consider
only the trajectories on the slow timescale.
In this section we suggest
an experimental observable that captures the trace of
the anomalous behavior of the fitness flux in the microevolutionary limit.
When the additional adaptation is taking place, the following observable:
\begin{equation}\label{eq:obs}
  A_t(\tau) = \frac{1}{|\tau|} \langle g_t^{-1} \left( X_{t+\tau} - X_t \right)^2 \rangle=
  \frac{1}{|\tau|} \left\langle \frac{ \left( X_{t+\tau} - X_t \right)^2}{ X_t (1 - X_t)}\right \rangle
\end{equation}
 presents a linear 
growth for short times (As shown in figure 2).
This follows form the fact that such quantity is closely related to the fitness flux.
Indeed, when
mutations are negligible compared with the selection and the population is large, the slope of $A(\tau)$ is equal to fitness flux rate
(see the supplemental material).
Therefore computing such observable allows to discriminate
the dynamic balance of adaptation and fitness loss
from the static, non-adapting case.
Note that as for the fitness flux once the separation
of time scale is large enough to allow the asymptotic expansion
the additional contribution is independent of the specific
environmental correlation time.
%

\section{Example}
To illustrate our findings let us now consider a frequency independent selection 
with $s(Y_t) = \sigma Y_t$ so that the global fitness coincides with the average fitness (up to a constant).
When $Y_t>0$ genotype $A$ is favored by selection whereas
$Y_t<0$ one favors $B$. 
The fitness of the population, then reads 
\begin{equation}
 F(x,y) = \sigma x y=f_A(y)x+f_B(y)(1-x)+C_0=\langle f \rangle+C_0
\end{equation}
 where $C_0=-f_B$ is a  term constant in $x$.

For example we take an environment dynamics
described by
 an Ornstein-Uhlenbeck process with constant diffusion $D$ and spring constant $K$ (of order $1$)
\begin{equation}\label{eq:env_ex}
 dY_t = -\epsilon^{-1}KY_t dt + \epsilon^{-1/2}\sqrt{2D} \cdot dB_t
\end{equation}
where $dB_t$ is a Wiener process and $\epsilon\ll 1$ is the parameter accounting for the timescale separation between the
environment and the genotype evolution.
At the steady state
$\overline{y} = 0$ and $\overline{y^2} = D/K$.
This implies that  $\overline{s}=0$ and that, consequently, the effective evolution on the slow timescales seems neutral.
Indeed the regular term of the fitness flux $\langle \phi_{\mathit{eff}} \rangle$
is equal to zero. However, on the timescale of the environmental variation adaptation is taking place and this is
accounted for by the additional contribution we have derived which in the present case gives
%
\begin{equation}\label{eq:eff_df_ex}
     d \langle F(X_t,Y_t) \rangle = \underbrace{\frac{D}{K} \sigma^2 \langle g \rangle }_{\langle d\phi(t)\rangle}
  \underbrace{-\frac{D}{K} \sigma^2 \langle g \rangle}_{\langle
  dR\rangle } = 0
  \end{equation}
showing that the constant fitness  average arises from the balance of two finite terms.

\section{Conclusions and discussion}
Population genetics under varying or fluctuating selection
has been considered in a number of studies (see for instance Refs. \cite{mustonen08,mustonen10,peischl12,otinowski11,tanasenicola12,bell10,rivoire14,held14}).
In particular, the case when the selection that acts on a population changes on timescales different from the ones
of the evolutionary process was also considered in Ref.~\cite{tanasenicola12}. There, the authors analyzed a scenario specular to the one we have studied.
They focused on a very slowly changing environment and found that such slow variation impacts
on the fitness of the population providing a correction term of the order of the environment change rate.
Conversely, we have dealt with the opposite limit and
shown that when a population evolves in a rapidly changing environment its dynamics is governed
by the average of the selection coefficients experienced across the various conditions. If the environment 
reaches an equilibrium the global fitness 
of the population is then stationary. Such stationarity emerges from a tug of war between the loss of fitness
due to the fast environmental changes and a continuous adaptation taking place at the fast timescales of
the environmental variations. On average the two contributions
scale linearly in time but they exactly compensate yielding a constant global fitness.
They do not depend on the magnitude of timescale separation (provided it is large enough to allow 
the asymptotic expansion).
Our findings support the idea that in varying environments a constant population fitness
does not imply that no adaptation is taking place. To further investigate the issue we have 
suggested an experimental observable based on genotypic frequencies alone that is able to discriminate
the case of adaptive behavior at constant fitness from the static
non adaptive one.
In closing, we note that similar reasonings have been adopted in more complex and exhaustive studies
attempting to reconcile the data about the rapid evolution of extant populations with the 
stasis over long time periods emerging from fossil records \cite{Eldredge2005}.
However, more complete approaches than the one presented here are needed to settle the issue.
Most importantly, one would need to identify a mechanism for the breaking of stasis and the speciation episodes
(which is not present in our analysis) and to extend the formalism
adopted in this paper to account 
for the influence of geographical effects on evolution.
\section{Supplemental material}
Here we provide the general derivation of the results discussed in the main text.
\subsection{Genotype dynamics}

The Kimura-Ohta equation (2) in the main text  describes the evolution of two genotypes
 in a population under mutations,
 random genetic drift, and natural selection.
For the sake of generality we consider here the case
of a haploid population with $n$ possible genotypes, referring
 to the supporting info of Ref.~\cite{mustonen10}. 
A generic genotype frequency is $x^\alpha$, where $\alpha = 1, \ldots n$, 
such that $x^\alpha \geq 0$ and $\sum_{\alpha = 1}^{n} x^\alpha = 1$.
 Therefore the dynamics is defined in the space of the $n-1$ independent genotype frequencies.

We will use the implicit summation over the $n-1$ independent frequencies when the same coordinate 
appears in a product as upper and lower index. For example: $m_\alpha x^\alpha = \sum_{\alpha = 1}^{n-1} m_\alpha x^\alpha$.

We generalize the quantities that we have seen for the two-genotypes equation as follows:
\begin{itemize}
\item response coefficient:
\begin{equation}
  g^{\alpha \beta}(x) = 
  \begin{cases}
     -x^\alpha x^\beta         & \text{if $\alpha \neq \beta$} \\
     x^\alpha (1 - x^\alpha)   & \text{if $\alpha = \beta$}
  \end{cases}
\end{equation}
\item selection coefficient:
\begin{equation}
  s_\alpha(x,t) = f_\alpha(x,t) - f_n(x,t)
\end{equation}
where $f_\alpha(x,t)$ is the fitness of the genotype $\alpha$,
\item mutation coefficient:
\begin{equation}
m_\alpha(x) = \hat{\mu}^\alpha_\beta x^\beta + \mu_n^\alpha
\end{equation}
where we use the implicit summation over $\beta$, $\mu_\alpha^\beta$ is 
the mutation rate (probability of mutation per generation) from the
 genotype $\alpha$ to $\beta$ and $\hat{\mu}_\beta^\alpha$ is:
\begin{equation}
  \hat{\mu}_\beta^\alpha = 
  \begin{cases}
     \mu_\beta^\alpha - \mu_n^\alpha                         & \text{if $\alpha \neq \beta$} \\
     -\sum_{\gamma = 1}^n \mu_\alpha^\gamma - \mu_n^\alpha   & \text{if $\alpha = \beta$}
  \end{cases}
\end{equation}
\end{itemize}

Then the general Kimura-Ohta equation
 describing the dynamics of $P(\bm{x},t)$, the probability of a specific configuration of genotype
 frequencies $\bm{x} = (x^1, \ldots, x^{n-1})$ at time $t$, reads as follows:
\begin{equation}\label{eq:ko_mult}
\frac{\partial}{\partial t} P =
 - \frac{\partial}{\partial x^\alpha} \biggr( \left( s_\beta g^{\alpha\beta}(x) + m^\alpha \right) P \biggr) 
+ \frac{1}{2N} \frac{\partial^2}{\partial x^\alpha x^\beta} \left( g^{\alpha\beta} P \right)
\end{equation}
where $t$ labels the time in generation units.
Note that this formalism can be applied also to the study of allele frequencies instead of genotype ones. 
The Kimura-Ohta equation would then  describe the temporal evolution of the $n$ allele frequencies within a population.
Of course we would have to replace the mutation rates among genotypes with the mutation rates among alleles
 and the fitness has to refer to the alleles.

{\it Fluctuating environment.}
In the main text we have discussed the case when the environment, in which the population
evolves, rapidly
changes in time according to a stochastic process following an
Ornstein-Uhlenbeck dynamics.
Here we consider a more general process with the sole assumption that
the environment dynamics does admit a stationary equilibrium.
With the results of Appendix A 
we can express
the general Kimura-Ohta equation (\ref{eq:ko_mult})
as a stochastic differential equation (SDE) for the $n-1$ independent genotype frequencies.
 By denoting $s^\alpha=s_\beta g^{\alpha\beta}$ the coupled SDEs  then read
\begin{equation}\label{eq:ko_sde_gen}
\begin{cases}
dX^\alpha_t = \left(s^\alpha(\bm{X}_t,Y_t) + m^\alpha(\bm{X}_t)\right) dt + d^{\alpha\beta}(\bm{X}_t) \cdot dW_t^\beta
\\
dY_t = \epsilon^{-1}a(Y_t,t) dt + \epsilon^{-1/2}\sqrt{2D(Y_t,t)} \cdot dB_t
\end{cases}
\end{equation}
with 
\begin{equation}
d^{\alpha\gamma}d^{\gamma\beta} = \frac{1}{N} g^{\alpha\beta}
\end{equation}
where $\alpha,\beta=1,\ldots,n-1$ label the genotypes and $\cdot$ denotes the It\=o product.
 $dB_t$ and $dW_t^\alpha$ are independent
 Wiener processes:
 $\langle dB_t \rangle = \langle dW_t^\alpha \rangle =\langle dB_{t'}\cdot dW_t^\alpha \rangle= 0$ $\forall \alpha$, 
$\langle dW_t^\alpha\cdot dW_{t'}^\beta \rangle = dt\delta(t-t') \delta_{\alpha \beta}$ and 
$\langle dB_t\cdot dB_{t'} \rangle = \delta(t-t')dt$.
This system of SDEs corresponds to the Fokker-Planck equation for the propagator
 $p(\bm{x},y,t|\bm{x}_0',y_0,t_0)$, which is the joint probability distribution for the 
variables $\bm{x}$, $y$ at time $t$, given the initial condition $\bm{x}_0$, $y_0$. 
Making use of the results shown in the appendix A 
we obtain
\begin{equation}\label{eq:fp}
\partial_{t} p = - \partial_{\alpha} ((m^{\alpha}(\bm{x}) + s^{\alpha}(\bm{x},y))p)
 + \frac{1}{2N} \partial _\alpha \partial_\beta (g^{\alpha\beta}(\bm{x})p) 
+ \epsilon^{-1} \left[ -\partial_y(a(y,t)p) + \partial^2_y(D(y,t)p)\right]
\end{equation}

{\it Relevant timescales.}
Let us briefly consider what are the timescales involved in the
evolution process we are considering.
The basic unit of time we have  chosen is that of a generation.
To explicitly account for the fact that the environment is a fast process we have introduced the
parameter $ \epsilon\ll1 $.
By doing so, we can identify the typical environmental time scale with $\theta = \epsilon^{-1} t$.
The evolutionary forces act on the genotype frequency with the following timescales: 
\begin{itemize}
\item mutations lead to significant changes after about: $\tau_m = m(x)^{-1}$
generations.
\item The time scale of the natural selection is: $\tau_s = g(x)^{-1}s(x,y)^{-1}$ generations.
\item The random genetic drift has a typical time of $\tau_{rgd} = g(x)^{-1}N$ generations.
\end{itemize}
Assuming that the trajectory spends most of the time away from the fixation ($x = 0$ or $x = 1$) and ignoring multiplicative coefficients,
we can conclude that the underlining characteristic times of genotypes evolution expressed in generation units are:
\begin{equation}\label{eq:t_gen}
\tau_m \simeq \mu^{-1} \hspace{2cm} \tau_s \simeq s(x,y)^{-1} \hspace{2cm} \tau_{rgd} \simeq N
\end{equation}

The typical time of environmental changes are the following:
\begin{equation}\label{eq:t_env}
\tau_e^{(a)} \simeq \epsilon a^{-1} \hspace{2cm} \tau_e^{(D)} \simeq \epsilon D^{-1}
\end{equation}

The micro-evolutionary regime refers to the scenario for which
the characteristic times of environmental changes (\ref{eq:t_env}) 
are much shorter than the ones of genotype variations (\ref{eq:t_gen}). 
If we consider the coefficients $a$ and $D$ of the same order of the genotype dynamics ones,
the micro-evolutionary limit  corresponds simply to take the limit  $\epsilon \rightarrow 0$. 
For the reminder of this section, we will refer to this instance.

It is important to note that the system loses physical meaning when 
the typical time of environmental changes is comparable to or lesser than a generation.
 We therefore have to consider that $T_e^{(a,D)} \gg 1$ must be satisfied.

{\it Effective genotype dynamics in the micro-evolutionary limit.}

 As mentioned above,  the time scale of the genotype evolution is defined by $t = O(1)$ while
 the environment evolves with a time scale equal to $\theta = \epsilon^{-1}t$.
Micro-evolution corresponds to the limit of $\epsilon \rightarrow 0$
which means that the two timescales are well separated.
We intend to derive the effective genotype dynamics taking place
at time $t = O(1)$ by means of asymptotic techniques (see e.g.~\cite{pavliotis}). 
The separation of
 timescales allows us to write:
\begin{equation}
\partial_t \rightarrow \epsilon^{-1}\partial_\theta + \partial_t
\end{equation}
and suggests to expand the propagator as
\begin{equation}
p = p^{(0)} + \epsilon p^{(1)} + O(\epsilon^2)\;.
\end{equation}
{\it Order $\epsilon^{-1}$.---}
Substituting these identities in the Fokker-Planck (\ref{eq:fp}), at the lowest order $\epsilon^{-1}$, we have:
\begin{equation}
\partial_{\theta} p^{(0)} = -\partial_y(a(y,t)p^{(0)}) + \partial^2_y(D(y,t)p^{(0)})\;.
\end{equation}
This equation describes the dynamics at the fast time scale of environmental evolution.
 Indeed, the terms dependent on genotype frequencies do not appear.
The fast environmental dynamics relaxes instantly to its 
equilibrium distribution $w(y)$, which satisfies the following relations:
\begin{equation}
-\partial_y(a(y,t)w(y)) + \partial^2_y(D(y,t)w(y)) = 0
\end{equation}
\begin{equation}
\int dy \hspace{0.1cm} w(y) = 1\;.
\end{equation}

Since the relaxation of the environment to the equilibrium distribution is very fast
 (of order of $\epsilon$) with respect to the time scale of genotypes evolution, we
 can write down the general solution for the propagator at the lowest order in the following way:
\begin{equation}
p^{(0)}(\bm{x},y,t) = \xi(\bm{x},t) w(y)
\end{equation}
where $\xi(\bm{x},t)$ describes the slow dynamics of the genotype frequencies. Note that,
 at the lowest order, the two variables $\bm{x}$ and $y$ are independent.

{\it Order $\epsilon^{0}$.---}
The Fokker Planck equation at order $\epsilon^0$ is
\begin{eqnarray}\label{eq:fp_0}
&&\left( \partial_\theta + \partial_y a(y,t) - \partial^2_y D(y,t) \right)p^{(1)} 
= \\\nonumber
&&- \partial_{t} p^{(0)} - \partial_{\alpha}\left[\left(m^{\alpha}(\bm{x}) + s^{\alpha}(\bm{x},y)\right)p^{(0)}\right] 
+ \frac{1}{2N} \partial _\alpha \partial_\beta \left[g^{\alpha\beta}(\bm{x})p^{(0)}\right]\;.
\end{eqnarray}
Integrating both terms of the equation in $dy$, one can obtain the solvability condition for this equation:
\begin{equation*}
\int \left\{ - \partial_{t} \xi(\bm{x},t)  - \partial_{\alpha}\left[\left(m^{\alpha}(\bm{x})
 + s^{\alpha}(\bm{x},y)\right)\xi(\bm{x},t)\right] 
 + \frac{1}{2N} \partial _\alpha \partial_\beta \left[g^{\alpha\beta}(\bm{x})\xi(\bm{x},t) \right] \right\}w(y)\,dy  = 0
\end{equation*}
and, using the normalization of the equilibrium distribution, one finds  the effective Kimura-Ohta equation:
\begin{equation}\label{eq:eff_ko}
\partial_{t} \xi = - \partial_{\alpha} \left[(m^{\alpha}(\bm{x}) 
+ \overline {s^{\alpha}(\bm{x})})\xi \right] 
+ \frac{1}{2N} \partial _\alpha \partial_\beta \left[g^{\alpha\beta}(\bm{x})\xi \right]
\end{equation}
that describes the effective dynamics of the genotype frequencies in the limit $\epsilon \rightarrow 0$.
 Note that this is simply
equation (\ref{eq:ko_mult}) with the selection coefficient substituted by an effective one
 which is independent of the environmental state:

\begin{equation}
\overline {s^{\alpha}(\bm{x})} = \int   s^{\alpha}(\bm{x}, y) w(y)\,dy\;.
\end{equation}

\subsection{Irreversibility and adaptation: the fitness flux}

As discussed in the main text, the concept of fitness flux
introduced in Ref.~\cite{mustonen10} provides a clear connection
between adaptation and irreversibility in the stochastic evolution of
a population. Namely, the fitness flux is equivalent
to the entropy produced in the environment by the stochastic
process describing the genotype evolution.
For a stochastic system the entropy produced along a trajectory
is defined as the log-ratio of the probability of observing the trajectory
to the one of observing the its time reversal \cite{chetrite08}:
\begin{equation}\label{eq:entro}
S\left(\bm{x}_{[0,t]}\right) = \log\frac{\mathcal{P}\left(\bm{x}_{[0,t]}\right)}{\mathcal{P}\left(\bm{x}^T_{[t,0]}\right)}
\end{equation}
where $\bm{x}^T_{[t,0]}$ is the reverse history 
of $\bm{x}_{[0,t]}$. This ratio measures the irreversibility of a certain path so that if the trajectory is time reversible 
i.e. ${\mathcal{P}\left(\bm{x}_{[0,t]}\right)}={\mathcal{P}\left(\bm{x}^T_{[t,0]}\right)}$ there is
no entropy production $S
=0$.
Jarzynski equality states that
\begin{equation}
\langle e^{-S\left(\bm{x}_{[0,t]}\right)}\rangle = 1
\end{equation}
 which implies via Jensen inequality $\langle S_t\rangle\ge0$
where $\langle \rangle$ denotes the average over trajectories%
.
 For the genotype dynamics described by eq.~(\ref{eq:ko_sde_gen}), once the environment has reached
its equilibrium state, entropy production reads~\cite{chetrite08,mustonen10}:
 \begin{equation}
S = 2N \int_{t_0}^t \hspace{1mm} \underbrace{s_{\alpha}(\bm{X}_\tau,Y_\tau) \circ dX^\alpha_\tau}_{d\phi(\tau) } - \Delta H
\end{equation}
where $\Delta H(\bm{x}) = H(t_f,x_f) - H(t_0,x_0)$ and
 $H(x,t) = \log\left( \frac{P(x,t)}{P_0(x)} \right)$. $P_0(x)$ is the 
equilibrium distribution given by the mutation-drift process in the absence of natural selection.
If we define the cumulative fitness flux of the trajectory $\bm{x}_{[0,t]}$ as 
$\Phi(\bm{x})=\int_{0}^t s_{\alpha}(\bm{X_\tau},Y_\tau) \circ dX^\alpha_\tau$ 
we then readily have:
\begin{equation}
\frac{\mathcal{P}(\bm{x}^T)}{\mathcal{P}(\bm{x})} = e^{- 2N \Phi (\bm{x}) + \Delta H(\bm{x})}
\end{equation}
and from the Jarzynski equality we can derive the fitness flux theorem:
\begin{equation}
\left\langle e^{- 2N \Phi (\bm{x}) + \Delta H(\bm{x})} \right\rangle = 1\;.
\end{equation}
An immediate consequence is the inequality (obtained with the Jensen's inequality):
\begin{equation}
2N \langle \Phi \rangle \geq \langle \Delta H \rangle
\end{equation} 
which gives a lower bound for the  mean fitness flux for every evolutionary 
process described by the Kimura-Ohta equation. 
Notably, this relation becomes an identity when the evolutionary process is in a state of equilibrium.
 In this context, the evolution obeys the local detailed balance, which implies the reversibility of each trajectory:
\begin{equation}
\mathcal{P}(\bm{x}^T) = \mathcal{P}(\bm{x}) \hspace{1cm} \Longrightarrow \hspace{1cm} 2N \Phi(\bm{x}) = \Delta H(\bm{x})\;.
\end{equation}
In a stationary state the
genotype distribution is constant and the average boundary vanishes
so that we have
\begin{equation}
\langle d\phi \rangle \ge 0\;.
\end{equation}
This relation shows that at the steady state adaptation always occurs
 and the cumulative fitness flux always increases, leading to the irreversibility of the evolutionary process.

{\it Fitness flux in the micro-evolutionary limit.}

Let us now consider the micro-evolutionary limit of the fitness flux as done in
the previous section for the genotype dynamics ($\epsilon \rightarrow 0$).
The fitness flux in an infinitesimal interval of time is given by:
\begin{equation}
d \phi_\tau = s_{\alpha}(\bm{X_\tau},Y_\tau) \circ dX^\alpha_\tau
\end{equation}
This equation defines the stochastic process for the cumulative fitness flux, such that $\Phi = \int_{t_0}^t d \phi_\tau$.

First of all it is useful to express $d\phi_\tau$ with an It\=o product instead of a Stratonovich one:
\begin{equation}
d \phi_\tau = \left( s_\alpha g^{\alpha\beta}(s_\beta + m_\beta) + \frac{g^{\alpha\beta}}{2N} \frac{\partial s_\alpha}{\partial x^\beta} \right) dt + s_\alpha d^{\alpha\beta} \cdot dB_\tau^\beta + o(dt)
\end{equation}

The system involving this equation and 
the ones in~(\ref{eq:ko_sde_gen}) is equivalent to the following Fokker-Planck equation for the propagator
 $q(\bm{x},y,\phi,t|\bm{x}_0,y_0,\phi_0,t_0)$
\begin{eqnarray}\label{eq:fp_ff}
\frac{\partial q}{\partial t} =&& 
\epsilon^{-1} \left( - \frac{\partial}{\partial y}\left(a \hspace{1mm} q \right) +
  \frac{\partial^2}{\partial y^2}\left(D \hspace{1mm} q \right) \right) - 
\frac{\partial}{\partial x^{\alpha}} \left(g^{\alpha\beta}(s_\beta + m_\beta)q \right) + 
\\\nonumber &&
+ \frac{1}{2N} \frac{\partial^2}{\partial x^{\alpha} \partial x^{\beta}} \left(g^{\alpha\beta}q \right) 
- \frac{\partial}{\partial \phi} \left( \left( s_\alpha g^{\alpha\beta}(s_\beta + m_\beta) + \frac{g^{\alpha\beta}}{2N} \frac{\partial s_\alpha}{\partial x^\beta} \right) q \right) +
\\\nonumber&&
+  \frac{1}{2N} \frac{\partial^2}{\partial \phi^2} \left(s_\alpha s_\beta g^{\alpha\beta} q \right) +
 \frac{1}{N} \frac{\partial^2}{\partial \phi \partial x^\alpha} \left(s_\beta g^{\alpha\beta} q \right)
\end{eqnarray}
which can be verified using the results in the appendix  A.\\

{\it Multi-scale derivation.}\\
As for the genotype dynamics, we take a multi-scale approach.
The time scale of genotype evolution is defined by $t = O(1)$, while environment 
evolves with time $\theta = \epsilon^{-1}t$. Using the time scales separation, we can write down:
\begin{equation}
\partial_t \rightarrow \epsilon^{-1}\partial_\theta + \partial_t
\end{equation}
\begin{equation}
q = q^{(0)} + \epsilon q^{(1)} + O(\epsilon^2)
\end{equation}

{\it Order $\epsilon^{-1}$.---}

The Fokker Planck equation~(\ref{eq:fp_ff}), at order $\epsilon^{-1}$, reads:
\begin{equation}
\partial_{\theta} q^{(0)} = -\partial_y(a(y,t)q^{(0)}) + \partial^2_y(D(y,t)q^{(0)})\;.
\end{equation}

Considering that the environment dynamics rapidly relaxes to its equilibrium distribution $w(y)$:
\begin{equation}
-\partial_y(a(y,t)w(y)) + \partial^2_y(D(y,t)w(y)) = 0
\end{equation}
\begin{equation}
\int dy \hspace{0.1cm} w(y) = 1
\end{equation}
the general solution for the propagator, at the lowest order, is then:
\begin{equation}
q^{(0)}(x,y,\phi,t) = \chi(x,\phi,t) w(y)
\end{equation}
where $\chi(x,\phi,t)$ describes the slow dynamics of the genotype frequencies and the fitness flux in the micro-evolutionary limit.

{\it Order $\epsilon^0$.---}

At order $\epsilon^0$, Eq.~(\ref{eq:fp_ff}) reads:
\begin{equation}
\begin{split}
\left( \frac{\partial}{\partial \theta} + 
\frac{\partial}{\partial y} a - \frac{\partial^2}{\partial y^2}
 D \right)  q^{(1)} = - \frac{\partial q^{(0)}}{\partial t} - \frac{\partial}{\partial x^{\alpha}}
 (g^{\alpha\beta}(s_\beta + m_\beta)q^{(0)}) + 
\\
+ \frac{1}{2N} \frac{\partial^2}{\partial x^{\alpha} \partial x^{\beta}} (g^{\alpha\beta}q^{(0)}) 
- \frac{\partial}{\partial \phi} \left( \left( s_\alpha g^{\alpha\beta}(s_\beta + m_\beta) 
+ \frac{g^{\alpha\beta}}{2N} \frac{\partial s_\alpha}{\partial x^\beta} \right) q^{(0)} \right) + 
\\
+ \frac{1}{2N} \frac{\partial^2}{\partial \phi^2} (s_\alpha s_\beta g^{\alpha\beta} q^{(0)}) 
+ \frac{1}{N} \frac{\partial^2}{\partial \phi \partial x^\alpha} (s_\beta g^{\alpha\beta} q^{(0)})
\end{split}
\end{equation}

Then, we impose the solvability condition:
\begin{equation*}
\begin{split}
\int dy \left( - \frac{\partial q^{(0)}}{\partial t} 
- \frac{\partial}{\partial x^{\alpha}} \left(g^{\alpha\beta}(s_\beta + m_\beta)q^{(0)}\right) +
 \frac{1}{2N} \frac{\partial^2}{\partial x^{\alpha} \partial x^{\beta}} \left(g^{\alpha\beta}q^{(0)}\right) \right.
\\
\left. - \frac{\partial}{\partial \phi}
 \left( \left( s_\alpha g^{\alpha\beta}(s_\beta + m_\beta) +
 \frac{g^{\alpha\beta}}{2N} \frac{\partial s_\alpha}{\partial x^\beta} \right) q^{(0)} \right) + \right. 
\\
\left. + \frac{1}{2N} \frac{\partial^2}{\partial \phi^2}
 \left(s_\alpha s_\beta g^{\alpha\beta} q^{(0)}\right) + 
\frac{1}{N} \frac{\partial^2}{\partial \phi \partial x^\alpha} \left(s_\beta g^{\alpha\beta} q^{(0)}\right) \right) = 0  
\end{split}
\end{equation*}

Using $q^{(0)} = \chi \hspace{1mm} w$ and the normalization of $w(y)$, one obtains the effective dynamics at the lowest order:
\begin{equation}\label{eq:eff_fp_ff}
\begin{split}
\frac{\partial \chi}{\partial t} = - \frac{\partial}{\partial x^{\alpha}} \left(g^{\alpha\beta}(\overline{s_\beta} 
+ m_\beta)\chi\right) + \frac{1}{2N} \frac{\partial^2}{\partial x^{\alpha} \partial x^{\beta}} 
\left( g^{\alpha\beta} \chi \right) - 
\\
- \frac{\partial}{\partial \phi} \left( \left( \overline{s_\alpha s_\beta} g^{\alpha\beta} +
 \overline{s_\alpha} g^{\alpha\beta} m_\beta + \frac{g^{\alpha\beta}}{2N} \overline{\frac{\partial s_\alpha}{\partial x^\beta}} \right) \chi \right) &
\\
 +  \frac{1}{2N} \frac{\partial^2}{\partial \phi^2} \left(\overline{s_\alpha s_\beta} g^{\alpha\beta} \chi \right) 
+ \frac{1}{N} \frac{\partial^2}{\partial \phi \partial x^\alpha} \left(\overline{s_\beta}  g^{\alpha\beta} \chi \right) &
\end{split}
\end{equation}

where $\overline{f}$ denotes the average over the equilibrium distribution of the environment:
\begin{equation}
\overline{f(x,t)} = \int dy \hspace{1mm} f(x,y,t) w(y)\;.
\end{equation}

{\it Effective fitness flux}\\
The effective Fokker Planck equation (\ref{eq:eff_fp_ff}) on the timescale $t = O(1)$ 
can be expressed as a system of SDEs (using the results in the appendix A. 
The SDE for the variable $X_t$ of course corresponds to the effective Kimura-Ohta equation (\ref{eq:eff_ko}):
\begin{equation}
dX^\alpha_t = g^{\alpha\beta}(\overline{s_\beta} + m_\beta) dt + d^{\alpha\beta} \cdot dW_t^\beta\;.
\end{equation}
The fitness flux reads:
\begin{eqnarray}
d \phi_t &=& \left( g^{\alpha\beta}(\overline{s_\alpha s_\beta} -
 \overline{s_\alpha} \hspace{1mm} \overline{s_\beta}) +
 \frac{g^{\alpha\beta}}{2N} \overline{\frac{\partial s_\alpha}{\partial x^\beta}} \right) dt\\\nonumber
&+& \overline{s_\alpha} \cdot d X_t^\alpha + 
 \sqrt{\frac{g^{\alpha\beta}}{N} (\overline{s_\alpha s_\beta} - \overline{s_\alpha} \hspace{1mm} \overline{s_\beta})} \cdot dC_t
\end{eqnarray}
where $dC_t$ is an independent Wiener process, independent of $dW_t^\alpha$ and $dY_t$. Using a Stratonovich product, the expression becomes:
\begin{equation}
d \phi_t = \left( g^{\alpha\beta}(\overline{s_\alpha s_\beta} - \overline{s_\alpha} \hspace{1mm} \overline{s_\beta}) \right) dt + \overline{s_\alpha} \circ d X_t^\alpha + 
 \sqrt{ \frac{g^{\alpha\beta}}{N}(\overline{s_\alpha s_\beta} - \overline{s_\alpha} \hspace{1mm} \overline{s_\beta})} \cdot dC_t
\end{equation}

Now we can evaluate the cumulative fitness flux of the micro-evolutionary
 process in a finite interval of time ($\Phi = \int_{t_0}^t d \phi_\tau$)
\begin{eqnarray}
\lim_{\epsilon \rightarrow 0} \Phi &=& \underbrace{\int_{t_0}^t  g^{\alpha\beta}(\overline{s_\alpha s_\beta} 
- \overline{s_\alpha} \hspace{1mm} \overline{s_\beta}) dt + 
\int_{t_0}^t  \sqrt{ \frac{g^{\alpha\beta}}{N}(\overline{s_\alpha s_\beta} 
- \overline{s_\alpha} \hspace{1mm} \overline{s_\beta})} \cdot dC_t}_{\Phi_{anom}} \\\nonumber
&+& \underbrace{ \int_{t_0}^t \overline{s_\alpha} \circ dX_t^\alpha }_{\Phi_{eff}}
\end{eqnarray}
where $\Phi_{eff}$ stands for the effective cumulative fitness flux and represents the adaptation generated by 
the effective genotype dynamics (\ref{eq:eff_ko}). $\Phi_{anom}$ is the anomalous contribution.
If we consider the average anomalous cumulative fitness flux at the stationary state, the expression becomes:
\begin{equation}\label{eq:eff_ff_supp}
\lim_{\epsilon \rightarrow 0} \langle \Phi_{anom} \rangle = \langle g^{\alpha\beta}(\overline{s_\alpha s_\beta} - \overline{s_\alpha} \hspace{1mm} \overline{s_\beta}) \rangle (t-t_0)
\end{equation}
where we can see that it grows linearly in time.

An alternative derivation of the given results
can be obtained by adapting the findings of Ref.~\cite{bo14} concerning entropy production
for the current settings exploiting that
\begin{equation}
 \Phi(\bm{x})=\frac{ S-\Delta H}{2N}
\end{equation}
and that the boundary terms will not give an anomalous contribution.
By applying the definitions one
 finds that the anomalous contribution to the fitness flux is the following:
\begin{equation}
d {\phi}_{anom} = \frac{1}{2N}dS_{anom}=\frac{1}{2N}\left(\frac{1}{2}l(X_t) dt + \sqrt{l(X_t)} \cdot dC_t\right) 
\end{equation}
where, for the considered dynamics,
\begin{equation}
l(x) =4Ng^{\alpha\beta}(\overline{s_\alpha s_\beta}
- \overline{s_\alpha} \hspace{1mm} \overline{s_\beta})
\end{equation}
and $dC_t$ is a Wiener process.
\subsection{Fitness dynamics}
Let us now consider the steady state behavior of the global fitness.
From equation (4) in the main text we known that the infinitesimal variation of the global fitness is given by
\begin{equation}
d F(X_t,Y_t) = \underbrace{s(X_t,Y_t) \circ dX_t}_{d\phi(t)} + \underbrace{\frac{\partial F}{\partial y} \circ dY_t}_{dR(t)}\;.
\end{equation}
This relation describes how the fitness changes over time driven by two contributions:
 the first term is the fitness flux, which represents the variation of fitness at a fixed environmental state.
 As we discussed in the main text it pertains to the adaptation of the population to 
the environment. The second term,  $dR(t)$, describes the fitness variation due to environmental changes at a fixed genotype frequency.
At the stationary state we expect the average global fitness to be constant and indeed

\begin{equation}\label{eq:gf}
\begin{split}
& d \langle F(X_t,Y_t) \rangle = \langle F(X_{t+dt},Y_{t+dt}) \rangle - \langle F(X_t,Y_t) \rangle = 
\\
= \int dx \hspace{1mm} d & y  \hspace{1mm} F(x,y) p(x,y,t) - \int dx \hspace{1mm} dy \hspace{1mm} F(x,y) p(x,y,t+dt) = 0
\end{split}
\end{equation}
because the propagator $p(x,y,t)$, at the stationary state is constant in time. 
 Since the fitness flux is on average always greater than zero the environmental
 term should to lead to a compensating decrease of fitness:
\begin{equation}
\underbrace{\langle d\phi(t) \rangle}_{\ge 0} dt + \underbrace{\langle dR(t) \rangle}_{\le 0} dt = 0
\end{equation}

This relation tells us that a population is always subject to a loss of fitness (i.e. $\langle dR \rangle \le0$)
 because of a changing environment. In this context, a population must adapt itself continuously to the environment 
(i.e. $\langle d\phi \rangle \ge 0$) in order to keep a constant average fitness.
{\it Micro-evolutionary limit of global fitness.}
We know that in the micro-evolutionary limit the fitness flux dynamics displays an anomalous behavior.
We evaluate here the consequences of such anomaly on the global fitness.

At the stationary state, the regular term of the fitness flux $\langle \phi_{eff} \rangle$
is equal to zero because the system is at equilibrium.
This is due to the fact that the selection coefficient is by definition
the gradient of the global fitness and the effective selection coefficient is independent of the environment.
Thus, equation (\ref{eq:gf}) becomes:
\begin{equation}
\lim_{\epsilon \rightarrow 0} d \langle F \rangle = \langle d\phi_{anom} \rangle - \langle dR \rangle = 0
\end{equation}
and, using the explicit expression of the anomalous fitness flux (\ref{eq:eff_ff_supp}) 
we obtain that the average environment term is equal 
and opposite to the anomalous fitness flux.

This can be readily checked using the explicit expression 
of $dR$. 
For the sake of exemplification we consider the case in which the selection coefficient has
a simple linear dependence on the environment and does not depend on the genotype frequency:
\begin{equation}\label{eq:sel_coe_ex}
s_\alpha(x,y) = \sigma_\alpha y
\end{equation}
so that the global fitness up to an additive constant reads:
\begin{equation}
F(x,y) = \sigma_\alpha x^{\alpha} y\;.
\end{equation}
The differential of the mean fitness is then:
\begin{equation}
d \langle F(X_t,Y_t) \rangle = \underbrace{\sigma_\alpha \langle Y_t \circ dX^\alpha_t \rangle}_{\langle d \phi(t) \rangle dt} 
+ \underbrace{\sigma_\alpha \langle X^\alpha_t \circ dY_t \rangle}_{\langle d R(t) \rangle dt}
\end{equation} 

With the results of the previous section we can evaluate the average value of the fitness flux in microevolution,
which coincides its anomalous part as in eq.(\ref{eq:eff_ff_supp}).
In this example we consider the evolution of the environment (\ref{eq:ko_sde_gen})
to be an Ornstein-Uhlenbeck process with constant diffusion $D$ and spring constant $K$
\begin{equation}\label{eq:env_ex_supp}
 dY_t = -\epsilon^{-1}KY_t dt + \epsilon^{-1/2}\sqrt{2D} \cdot dB_t
\end{equation}
so that
$\overline{y} = 0$ and $\overline{y^2} = D/K$ just as Eq. (\ref{eq:env_ex}) in the main text. This corresponds to setting $a(y)=-Ky$ in (\ref{eq:ko_sde_gen}).
For the chosen selection coefficient (\ref{eq:sel_coe_ex}) the average fitness flux then reads:
\begin{equation}\label{eq:anom_ff}
lim_{\epsilon \rightarrow 0} \langle d\phi \rangle  
=\frac{D}{K} \sigma_\alpha \langle g^{\alpha\,\beta} \rangle \sigma_\beta\;.
\end{equation}

We are interested in evaluating the mean value of the
environmental term, which, consistently, must be equal to the opposite average fitness flux. 

It reads as follows:
\begin{equation}
\langle dR(t) \rangle  = \sigma_\alpha \langle X^{\alpha}_t \circ {dY_t} \rangle = - \epsilon^{-1} K\sigma_{\alpha} \langle X^{\alpha}_t Y_t \rangle dt
\end{equation}
where we have substituted $dY_t$ with its explicit expression (\ref{eq:env_ex_supp}) and already put to zero the noise term
which does not contribute to the average.

Let us now focus on the quantity $\langle X^\alpha_t Y_t \rangle$. Its explicit form in microevolution reads:
\begin{equation}
\langle X^\alpha_t Y_t \rangle = \int d\bm{x} \hspace{1mm} dy \hspace{1mm} \left( p^{(0)} + \epsilon p^{(1)} + O(\epsilon^2) \right) x^\alpha y
\end{equation}
where we have used the perturbation series of the propagator in $\epsilon$.

In this example we can obtain implicitly the expression of $p^{(0)}$ and $p^{(1)}$ 
using the multi scale method for the coupled system composed by the genotype evolution and the environmental dynamics.
Assuming that the environment has reached its equilibrium distribution, $w(y)$, the 
Fokker-Planck equation at order $\epsilon^{-1}$ provides an expression for the zero order propagator: 
\begin{equation}
p^{(0)} = w(y) \xi(\bm{x},t)
\end{equation}
and since $\overline{y}=0$ we see that the contribution at order $\epsilon^{0}$ vanishes.
Making use of Eq. (\ref{eq:fp_0}), we can derive the  solution for the first order in $\epsilon$ of the propagator.
Let us denote the operator 
$y\partial_{y}K + D \partial^2_{y}\equiv M^{\dagger}$ and 
$f(y)=- \partial_{t} p^{(0)} - \partial_{\alpha}\left[\left(m^{\alpha}(\bm{x}) + y\sigma_{\beta}g^{\alpha\beta}(\bm{x})\right)p^{(0)}\right] 
+ \frac{1}{2N} \partial _\alpha \partial_\beta \left[g^{\alpha\beta}(\bm{x})p^{(0)}\right]$ so that equation (\ref{eq:fp_0}) after the initial 
fast relaxation becomes:
\begin{equation}
 -M^{\dagger} p^{(1)}=f \,.
\end{equation}

 We then have that

 \begin{equation}
 p^{(1)} = - {M^\dagger}^{-1}f
 + \mbox{zero modes of $M^\dagger$}\;.
 \end{equation}
%
%
%
%
%
%
%

Note that $f$ can be split as follows:
 \begin{equation}
  f(y)=C_1 w(y)+C_2 yw(y)
 \end{equation}
with $C_1=- \partial_{t} \xi(\bm{x}) - \partial_{\alpha}m^{\alpha}(\bm{x}) \xi(\bm{x}) 
+ \frac{1}{2N} \partial _\alpha \partial_\beta \left[g^{\alpha\beta}(\bm{x})\xi(\bm{x})\right]$ and 
$C_2=-\sigma_{\beta}\partial_{\alpha}g^{\alpha\beta}\xi(\bm{x})$.
Then
\begin{equation}
 -\int dy\,y {M^\dagger}^{-1}f= -\int dy\,fM^{-1}y=+\frac{1}{K}\int dy\,fy=\frac{D}{K^2}C_2
\end{equation}
so that 

\begin{equation}
\int d\bm{x} \hspace{1mm} dy \hspace{1mm} \left( p^{(0)} + \epsilon p^{(1)} \right) x^\alpha y =
- \epsilon \frac{D}{K^2} \sigma_\beta \int d\bm{x} \hspace{1mm} x^\alpha \partial_\gamma(g^{\gamma \beta} \xi)  
\end{equation}
and, integrating by parts,
\begin{equation}
\langle X^\alpha_t Y_t \rangle= \epsilon \frac{D}{K^2}\sigma_\beta \langle g ^{\alpha \beta}\rangle 
\end{equation}
The expression of the response term, using (5.28), is then:
\begin{equation}
\langle dR(t) \rangle = -\frac{D}{K}\sigma_\alpha\sigma_\beta \langle g ^{\alpha \beta}\rangle dt
\end{equation}
which is equal and opposite to the anomalous fitness flux (\ref{eq:anom_ff}), showing that, at the stationary state, the global fitness is constant in time.
\subsection{Observable for the anomaly}
As shown before, when the environment changes very rapidly compared with the evolution of the genotype 
 frequencies, the genotype dynamics becomes independent of the environmental state.
In this regime, the dynamics is indistinguishable from the one of a 
population which evolves under an constant environment with selection coefficient equals to $\overline{s}$. 
However, in the case of a quickly changing environment, the population is adapting much more 
than a population under a non-fluctuating environment with apparently the same dynamics. This phenomenon 
is coded in the anomalous term of the fitness flux. 
We intend to find an observable which allows to discriminate these two cases.
Let us consider an example satisfying two simplifying conditions:
first
the contribution of mutations is negligible compared to the selection $g|s| \gg m$.
When the trajectory is close to a fixation state, the value of $g$ tends to
zero. Therefore, for the condition to hold we need to consider trajectories that are far from the fixation states. 
The second assumption is that the environmental fluctuations do not favor, on average, one specific
genotype, then the effective selection coefficient is zero:
$\overline{s} =0 
$.
Under these assumptions the average value of the fitness flux features only the anomalous term (\ref{eq:eff_ff_supp}) which reads
\begin{equation}
\langle d \phi \rangle =\langle g^{\alpha\beta}\overline{s_\alpha s_\beta}\rangle\;.
\end{equation}
We introduce the quantity:
\begin{equation}
O(\tau) = \frac{1}{|\tau|} \langle g_{\alpha \beta} ( X^\alpha_{t+\tau} - X^\alpha_t ) ( X^\beta_{t+\tau} - X^\beta_t) \rangle\;.
\end{equation}
As we can see in simulations (fig. 2 in the main text) 
this observable behaves differently if it is undergoing microevolution or in the case with a fixed neutral environment.
In particular, when the population evolves in a fluctuating environment, the 
observable shows a linear growth for $\tau \rightarrow 0$ ($\tau$ is smaller than the 
typical time of environmental evolution) 
\begin{equation}
\lim_{\tau \rightarrow 0^+} O(\tau) \simeq \langle g^{\alpha \beta} s_\alpha s_\beta \rangle \tau + n \frac{1}{N}
\end{equation}
This is the key feature of the anomalous adaptation, which is strictly related with the anomalous term of the fitness flux. 
If the environment is constant the observable is simply zero.

\subsection{Appendix A: From Wiener process to Fokker-Planck equation}\label{appendix:fp}

In this paragraph we study how to move from a system of coupled stochastic differential equations
 to the associated Fokker Planck equation.
Let us consider the system of It\=o stochastic differential equations of variables $X_t^\alpha$ and variables $Y_t^i$:
\begin{equation}
\begin{cases}
dX^\alpha_t = a(X_t,Y_t,t) dt + \sigma^{\alpha\beta}(X_t,Y_t,t) \cdot dW_t^\beta
\\
dY^i = b(X_t,Y_t,t) dt + \rho^{ij}(X_t,Y_t,t) \cdot dB_t^j + \psi^{i\alpha}(X_t,Y_t,t) \cdot dW_t^\alpha
\end{cases}
\end{equation}
with
\begin{equation}
\sigma^{\alpha\gamma} \sigma^{\gamma\beta} = 2 h^{\alpha\beta}\qquad\qquad \rho^{ik} \rho^{kj} = 2 r^{ij}
\end{equation}
where $\alpha,\beta,\ldots=1,\ldots,n_x$; $i,j,\ldots=1,\ldots,n_y$. $B_t^i$ and $W_t^\alpha$ are independent Wiener processes.
Note that we have assumed that the equation for $dY_t^i$ has a stochastic term ($\psi^{i\alpha}(X_t,Y_t,t) \cdot dW_t^\alpha$) 
dependent on the Wiener process in the equation for $X_t$.
Our aim is to deduce the Fokker Planck equation associated with this system, which describes the temporal evolution
 of the propagator $p(\bm{x},\bm{y},t|\bm{x}_0,\bm{y}_0,t_0)$.

Consider an arbitrary function defined along a trajectory $f(\bm{x},\bm{y},t)$, its average value at time $t$ is:
\begin{equation}\label{eq:app_ave}
\langle f(\bm{x},\bm{y},t) \rangle = \int d \bm{x} \hspace{0.1cm} d \bm{y} \hspace{0.1cm} f(\bm{x},\bm{y},t) p(\bm{x},\bm{y},t|\bm{x}_0,\bm{y}_0,t_0)
\end{equation}
then, we move from a continuous to a discrete-time description. The Wiener process is
 defined as: $\Delta W_t = \sqrt{\Delta t} \hspace{0.1cm}\omega_t$, where $\omega_t$ is a
 Gaussian variable with $\langle \omega_t \rangle = 0$ and $\langle \omega_t^2 \rangle = 1$.
The differential of the average value of $f$ is
\begin{eqnarray}
\Delta \langle f \rangle &=& \langle \frac{\partial f}{\partial t} \rangle \Delta t 
+  \langle \frac{\partial f}{\partial x^\alpha} \Delta x^\alpha \rangle + \langle \frac{\partial f}{\partial y^i} \Delta y^i \rangle +
\frac{1}{2}\langle \frac{\partial^2 f}{\partial x^\alpha \partial x^\beta} \Delta x^\alpha \Delta x^\beta \rangle \nonumber\\
 &+& \frac{1}{2} \langle \frac{\partial^2 f}{\partial y^i \partial y^j}
 \Delta y^i \Delta y^j \rangle + \langle \frac{\partial^2 f}{\partial x^\alpha \partial y^i} \Delta x^\alpha \Delta y^i \rangle\;.
\end{eqnarray}

Now we substitute the explicit expressions of the  processes $\Delta x^\alpha$ and $\Delta y^i$.
 For example, the second term on the RHS reads:
\begin{eqnarray}\label{eq:app_2term}
\langle \frac{\partial f}{\partial x^\alpha} \Delta x^\alpha \rangle  &=& \langle \frac{\partial f}{\partial x^\alpha}
 (a^\alpha(X_t,Y_t,t) \Delta t + \sigma^{\alpha\beta}(X_t,Y_t,t)\sqrt{\Delta t} \omega_t^\beta) \rangle\\\nonumber
 &=& \langle a^\alpha \frac{\partial f}{\partial x^\alpha} \rangle \Delta t + 
\langle \sigma^{\alpha\beta}\rangle \langle \omega_t^\beta \rangle \sqrt{\Delta t} =
 \langle a^\alpha \frac{\partial f}{\partial x^\alpha} \rangle \Delta t
\end{eqnarray}

where, thanks to the non-anticipative discretization, $\sigma^{\alpha\beta}(X_t,Y_t,t)$ is independent of $\omega_t^\beta$. 
After a few algebraic calculations, the final expression is:
\begin{eqnarray}
\Delta \langle f \rangle &=& \langle \frac{\partial f}{\partial t} 
+ a^\alpha \frac{\partial f}{\partial x^\alpha} + b^i \frac{\partial f}{\partial y^i} 
+ h^{\alpha\beta} \frac{\partial^2 f}{\partial x^\alpha \partial x^\beta}  \\\nonumber
&+&(r^{ij} + 
\frac{1}{2} \psi^{i\alpha}\psi^{j\alpha})\frac{\partial^2 f}{\partial y^i \partial y^j} + 
\sigma^{\alpha\beta} \psi^{i\beta} \frac{\partial^2 f}{\partial x^\alpha \partial y^i} \rangle \Delta t + o(\Delta t)\;.
\end{eqnarray}

Using the definition of the average value (\ref{eq:app_ave}), one has:
\begin{eqnarray}
\frac{d \langle f \rangle}{d t} = \langle \frac{\partial f}{\partial t} \rangle +
 \int d \bm{x} \hspace{0.1cm} d\bm{y} \hspace{0.1cm} f \frac{\partial p}{\partial t}
\end{eqnarray}
and, using (\ref{eq:app_2term}), one obtains:
\begin{eqnarray}
\frac{d \langle f \rangle}{d t} = \langle \frac{\partial f}{\partial t} \rangle +
 \int d \bm{x} \hspace{0.1cm} d\bm{y} \hspace{0.1cm}
 p \left( a^\alpha \frac{\partial f}{\partial x^\alpha} + b^i \frac{\partial f}{\partial y^i} +
 h^{\alpha\beta} \frac{\partial^2 f}{\partial x^\alpha \partial x^\beta} \right.\\ \nonumber
\left. + (r^{ij} + \frac{1}{2} \psi^{i\alpha}\psi^{j\alpha})\frac{\partial^2 f}{\partial y^i \partial y^j} +
 \sigma^{\alpha\beta} \psi^{i\beta} \frac{\partial^2 f}{\partial x^\alpha \partial y^i} \right)\;.
\end{eqnarray}

These two expressions imply that:
\begin{eqnarray}
\int d \bm{x} \hspace{0.1cm} d\bm{y} \hspace{0.1cm} f \frac{\partial p}{\partial t} 
- p \left( a^\alpha \frac{\partial f}{\partial x^\alpha} + b^i \frac{\partial f}{\partial y^i} 
+ h^{\alpha\beta} \frac{\partial^2 f}{\partial x^\alpha \partial x^\beta} \right. \\\nonumber
+\left.(r^{ij} + \frac{1}{2} \psi^{i\alpha}\psi^{j\alpha})\frac{\partial^2 f}{\partial y^i \partial y^j}
 + \sigma^{\alpha\beta} \psi^{i\beta} \frac{\partial^2 f}{\partial x^\alpha \partial y^i} \right) = 0
\end{eqnarray}
and, integrating by parts, the expression becomes (we choose the arbitrary function $f$ such that the boundary terms are zero):
\begin{eqnarray}
\int d \bm{x} \hspace{0.1cm} d\bm{y} \hspace{0.1cm} f \left( \frac{\partial p}{\partial t}
 + \frac{\partial}{\partial x^\alpha}(p a^\alpha) + \frac{\partial}{\partial y^i}(p  b^i) 
- \frac{\partial^2}{\partial x^\alpha \partial x^\beta}(p h^{\alpha\beta}) \right.\\\nonumber
-\left.\frac{\partial^2}{\partial y^i \partial y^j}(p(r^{ij}
+ \frac{1}{2} \psi^{i\alpha}\psi^{j\alpha}))
 - \frac{\partial^2}{\partial x^\alpha \partial y^i}(p \sigma^{\alpha\beta} \psi^{i\beta}) \right) = 0\;.
\end{eqnarray}

This expression implies that the integrand is zero and that finally leads to the Fokker
 Planck equation associated with the system (A.1), which describes the evolution of the 
propagator $p(\bm{x},\bm{y},t|\bm{x}_0,\bm{y}_0,t_0)$:
\begin{eqnarray}
\frac{\partial p}{\partial t}&=& - \frac{\partial}{\partial x^\alpha}(p a^\alpha)
 - \frac{\partial}{\partial y^i}(p  b^i)
  +\frac{\partial^2}{\partial x^\alpha \partial x^\beta}(p h^{\alpha\beta}) \\\nonumber
&+& \frac{\partial^2}{\partial y^i \partial y^j}(p(r^{ij} + \frac{1}{2} \psi^{i\alpha}\psi^{j\alpha}))
 + \frac{\partial^2}{\partial x^\alpha \partial y^i}(p \sigma^{\alpha\beta} \psi^{i\beta})\;.
\end{eqnarray}

\acknowledgements
SB acknowledges ICTP for hospitality.

\end{document}